\documentclass[conference]{IEEEtran}
\IEEEoverridecommandlockouts
\usepackage{cite}
\usepackage{amsmath,amssymb,amsfonts}
\usepackage{algorithmic}
\usepackage{graphicx}
\usepackage{textcomp}
\usepackage{xcolor}
\def\BibTeX{{\rm B\kern-.05em{\sc i\kern-.025em b}\kern-.08em
    T\kern-.1667em\lower.7ex\hbox{E}\kern-.125emX}}
\begin{document}

\title{Proof of Deep Learning: Approaches, Challenges, and Future Directions}

\author{\IEEEauthorblockN{Mahmoud Salhab}
\IEEEauthorblockA{\textit{Department of Computer Science and Mathematics} \\
\textit{Lebanese American University}\\
Beirut, Lebanon \\
mahmoud.salhab@lau.edu.lb}
 \and

\IEEEauthorblockN{Khaleel Mershad}
\IEEEauthorblockA{\textit{Department of Computer Science and Mathematics} \\
\textit{Lebanese American University}\\
Beirut, Lebanon \\
khaleel.mershad@lau.edu.lb}
}
\maketitle

\begin{abstract}
The rise of computational power has led to unprecedented performance gains for deep learning models. As more data becomes available and model architectures become more complex, the need for more computational power increases. On the other hand, since the introduction of Bitcoin as the first cryptocurrency and the establishment of the concept of blockchain as a distributed ledger, many variants and approaches have been proposed. However, many of them have one thing in common, which is the Proof of Work (PoW) consensus mechanism. PoW is mainly used to support the process of new block generation. While PoW has proven its robustness, its main drawback is that it requires a significant amount of processing power to maintain the security and integrity of the blockchain. This is due to applying brute force to solve a hashing puzzle. To utilize the computational power available in useful and meaningful work while keeping the blockchain secure, many techniques have been proposed, one of which is known as Proof of Deep Learning (PoDL). PoDL is a consensus mechanism that uses the process of training a deep learning model as proof of work to add new blocks to the blockchain. In this paper, we survey the various approaches for PoDL. We discuss the different types of PoDL algorithms, their advantages and disadvantages, and their potential applications. We also discuss the challenges of implementing PoDL and future research directions.
\end{abstract}


\section{Introduction}

Blockchain technology at its core is a type of distributed ledger that enables multiple users to achieve consensus without the need for a central authority. Additionally, it ensures the immutability of the stored records, thereby providing tamper-proofness. Bitcoin \cite{b0} is the most famous application that uses Proof of Work (PoW) as its consensus mechanism. The consensus algorithm is a crucial component of a blockchain network system. It helps make the system decentralized, transparent, auditable, secure, and tamper-resistant. The algorithm works by offering incentives to users who participate in the network, ensuring that everyone follows the rules and works towards the same goal \cite{survay-cons}.

Miners of Bitcoin consume a lot of computational resources due to the large amount of hash calculation required by PoW \cite{pow-analysis1, pos}. According to \cite{energy-consump} the annual electrical energy consumed by the Bitcoin network is estimated to be 87.2 TWh as of 2019, which is similar to the consumption of a country such as Belgium. To mitigate the amount of energy required, various solutions have been proposed, such as using an Application-Specific Integrated Circuit (ASIC) machine \cite{bitcoin-and-crypto}, using different consensus algorithms instead of PoW, such as Proof of Stake (PoS) \cite{pos}, Proof of Activity (PoA) \cite{poa0}, and Proof of Useful Work (PoUW) \cite{bitcoin-and-crypto}.

After the introduction of the PoUW concept, which aims to utilize computational power and wasted resources for solving hash puzzles in the PoW algorithm, researchers have been exploring ways to employ these resources to do useful things. For example, Primecoin \cite{pouw1} is a type of altcoin that utilizes a unique mining process where miners are required to find a specific prime number sequence known as Cunningham chains, as an alternative to the usual hash puzzles. While the discovery of Cunningham chains through this mining process holds mathematical and research significance, its practical applications in the real world remain uncertain \cite{b4}. Similarly, the Proof of Exercise (PoX) is a proposed mechanism outlined in \cite{pouw2} that falls under the PoUW category, requiring miners to perform specific exercises and provide proof of their outcomes. However, one significant limitation of the PoX mechanism is its reliance on an outsourced centralized board, which diminishes the decentralization feature of the blockchain.

To develop high-performance deep learning (DL) models, significant computational and memory resources are often required \cite{b9, dl-comp-power1}. To address this, researchers have proposed a novel approach that involves utilizing the computational power traditionally used for proof of work in blockchain to instead be used for training DL models. The supervised training of DL models is used to secure the blockchain in the Proof of Deep Learning \cite{b3}. However, the model proposed in \cite{b3} relies heavily on the integrity of the data provider. The authors in \cite{dl-bl-surv} surveyed the benefits of integrating deep learning and blockchain for data security, automatic decision-making, cumulative judgments, and enhanced robustness. They also touched on the concept of proof of deep learning but did not go into much detail.

In this study, we explore the nascent idea of using deep learning as a proof of work, which is still in its formative years and has a lot of potential for growth. Our research delves into the latest advances in this field and examines the various types of PoDL algorithms, including their benefits, drawbacks, and potential applications. Furthermore, we investigate the obstacles associated with implementing PoDL and identify potential areas for future research. To the best of our knowledge, this is the first paper that provides a comprehensive overview of the latest advances in PoDL. The paper is intended to be a valuable resource for researchers who are interested in learning more about this rapidly evolving field.

The paper is structured as follows: we start in section \ref{sec:background} by giving a high-level background of the blockchain and deep learning. In Section \ref{sec:podl}, we delve into the details of PoDL. Following this, Section \ref{sec:challanges} examines the challenges associated with PoDL systems. In Section \ref{sec:disc}, we discuss the current state and potential future directions of PoDL. Finally, we summarize our findings and conclude the paper in Section \ref{sec:conc}.

\section{Background}
\label{sec:background}
\subsection{Cryptocurrencies and Blockchain}

The blockchain is a distributed network of interconnected nodes that operate in a peer-to-peer manner, enabling the transfer of digital assets without the need for any intermediaries \cite{blck1}. Blockchain was originally built to support the work of the most famous cryptocurrency, Bitcoin \cite{blck2}.

Blockchain is a publicly distributed ledger that comprises a chain of blocks that store transactions. The chain expands continuously when a new block is added to it \cite{blck3}. All the transactions are processed in a decentralized way, which eliminates the need to have intermediaries to validate and verify them \cite{blck4}.

Decentralization, transparency, immutability, and auditability are among the essential features of blockchain technology, as outlined in \cite{blck5}.

In the Bitcoin blockchain, the system comprises a list of blocks where each block consists of three main items \cite{b0}:

\begin{itemize}
    \item Merkle Tree represents the transaction data.
    \item The previous block's cryptographic hash, except for the first block (the "Genesis Block") where it is hard coded.
    \item A nonce number used for consensus and validation.
\end{itemize}

For a blockchain to be considered valid, all of its blocks must be valid. A block, in turn, is considered valid if:

\begin{itemize}
    \item Each transaction in the block is valid.
    \item Contains the hash of the previous block (except for the first block).
    \item The hash of the entire block (the hash of the concatenation of the nonce, the hash of the previous block, and the Merkel tree) is less than a pre-defined value which is called the target difficulty. 
\end{itemize}

The target difficulty of a blockchain is frequently updated to maintain a consistent block generation rate (for example, 10 minutes in Bitcoin ) based on the network hash rate \cite{b0}.

Bitcoin leverages the Hashcash algorithm \cite{hashcash}, which entails the miners competing to validate a block by solving a hash puzzle. Miners compete to find the nonce value that meets the target difficulty. The hash function SHA-256 is utilized in Bitcoin due to its puzzle-friendliness property \cite{bitcoin-and-crypto}.

The process of hash puzzle solving which is known as Proof of Work (PoW) is expensive and requires substantial computational power as well as specialized hardware. All miners compete to propose the next block, and the first node that solves the puzzle will be compensated for its effort. The reward will be in the form of bitcoins, where the winner node receives two types of fees, which are \cite{incentive1}:

\begin{itemize}
    \item \textbf{Transaction fees}: These fees are paid by the creators of the block transactions to the miners.

    \item \textbf{Block reward}: The miner of the block mints a number of bitcoins, which decreases over time.
\end{itemize}

In the absence of a central node that everybody trusts to ensure that all nodes have the same ledger, there is a need for a process where nodes agree on the truth in such a trustless distributed network. In blockchain, this process is known as consensus, below are some of the conventional methods for achieving consensus in the blockchain:

\begin{itemize}

    \item \textbf{Proof of Work (PoW)}: Participants, also known as miners, will do an exhaustive search (i.e. brute force) to find the best nonce value that meets the target hash. Once it's found, the node that solved the puzzle will disseminate the block throughout the network, and then the block will be validated. Once it's validated, it will be added to the blockchain. The primary apprehension with the PoW approach is that miners are required to expend substantial resources in order to solve the puzzle and generate a block.

    \item \textbf{Proof of Stake (PoS)}: In this method, miners don't have to solve a mathematical puzzle, hence it's computationally efficient. The validation is done by selecting specific nodes for block validation. The chance of being selected is mainly correlated with the stake or the wealth of the node.

    \item \textbf{Delegated proof-of-stake (DPoS)}: In this elective consensus process, nodes that hold a stake in the network have the option to delegate transaction validation to another node through a voting process.

    \item \textbf{Byzantine Fault Tolerance}: Involves a group of nodes agreeing on a collective course of action such as validating a transaction, and was developed based on research on the Byzantine fault \cite{byzantine-fault}. 

    \item \textbf{Proof of Authority}: Similar to BFT, PoA involves delegating specific nodes in the network with authoritative control to achieve consensus based on majority votes when validating a block.

    \item \textbf{Proof of Elapsed Time}: Similar to PoA, PoET chooses a leader to generate new blocks in the blockchain by linking response time to a timer and selecting the node with the shortest expiration time as the leader.

    \item \textbf{Proof of Burn}: Requires validators to spend or burn their coins to create a new block and receive a reward. This process enhances the validator's stake in the network, while also reducing the number of coins in circulation and increasing the value of the remaining coins due to the coin-burning mechanism.

    \item \textbf{Proof of Importance}: Similar to PoA, this consensus mechanism selects validating nodes based on their stake in the network, but in this case, the stake is determined by their history of successful transactions and validations.

    \item \textbf{Proof of Capacity}: Also known as Proof of Space or Proof of Storage, it operates as an alternative to PoW by storing all possible nonce values on the hard drives of participating nodes.

\end{itemize}

\subsection{Deep Learning}

As a result of the increase in computational power and data accessibility, deep learning \cite{dl1} has experienced significant achievements across a range of application domains.

There are four categories into which various forms of deep learning can be grouped, as noted by \cite{dl-01}:

\begin{itemize}
    \item Deep Supervised Learning, which utilizes labeled data for training purposes.

    \item Deep Semi-supervised Learning, which operates on partially labeled data.
    
    \item Deep Unsupervised Learning, which does not rely on labeled data during the training process.

    \item Deep Reinforcement Learning, which is a technique used for learning in unfamiliar environments.
    
\end{itemize}

Given a model that maps input features X to target Y, where X represents the input data and Y is the target or the label of X, which can be continuous in the case of regression tasks and discrete in the case of classification tasks. Different optimization methods, such as SGD, Adagrad, AdaDelta, RMSprop, and Adam can be used to train the model on the given data \cite{optim}. These optimization techniques are used to minimize an objective function, which measures how good the model's predictions are. It is important to note that this function must be differentiable, as these optimization methods use gradients to minimize/maximize the objective function.

The training process is done by iteratively performing the below steps:

\begin{enumerate}
    \item The neural network processes input data, generates predictions and produces outputs by applying weights to the input and activating functions.

    \item The cost function evaluates the disparity between the predicted and actual outputs. This is used as a measure of how well the neural network is performing.
    
    \item Gradients are calculated using the cost function. These gradients are used to update the weights in the neural network.

    \item In the backward pass, the gradients calculated in step 3 are utilized for updating the weights of the neural network, which assists in enhancing the network's performance. This process is repeated many times until the network reaches an acceptable level of accuracy.
    
\end{enumerate}

Training a model on a large dataset can be challenging as it may not be feasible to pass the entire dataset at once. Therefore, the dataset is split into smaller batches that can be used iteratively to train the model. An epoch refers to the point where all batches have passed through the model once during the training process. Typically, model training involves multiple epochs rather than a single epoch.

In order to prevent overfitting where the model memorizes the training data, it is essential to split the data into training and testing sets. In practical scenarios, this data is further split into three sets: 
\begin{itemize}
    \item \textbf{Training set}: This is the data that the model will be trained on.
    \item \textbf{Validation set}: Used during the training process to select the parameters that lead to the best performance.
    \item \textbf{Test set}: Used to test the actual model performance.
\end{itemize}

The training and validation data are utilized to train the model and fine-tune the hyperparameters. On the other hand, the test data is leveraged to evaluate how the model performs on unseen data.

A diverse set of metrics are available for evaluating the efficacy of models, such as Accuracy, Precision, Recall, and F1-score, among others.

\section{Proof of Deep Learning (PoDL)}
\label{sec:podl}

In this section, we present a summary of how PoDL operates and the key participants involved in the PoDL-based network. We conducted a thorough investigation of existing research and examined the strategies employed to address the following aspects:

\begin{itemize}
    \item Approaches for rewarding honest mining participants.
    \item Methodologies for task initiation and model submission.
    \item Block validation and acceptance criteria, along with techniques for verifying the entire blockchain.
    \item Techniques employed to handle short model training time and enable incremental training across blocks.
    \item Approaches used to tackle challenges such as double spending, self-publishing, data privacy, and model size.
\end{itemize}

The original PoW technique requires nodes to compete to find out a nonce value in such a way that the resulting block hash meets certain criteria. This process is computationally expensive and does not utilize the computational resources for useful operations. To address this issue, PoDL was proposed. In PoDL, nodes compete to train deep learning models and use trained models as evidence of the work they've done. This results in utilizing the computational resources as well as making the blockchain robust against tampering by making it computationally infeasible for malicious nodes to tamper with any block.

\begin{figure*}
    \centering
    \includegraphics[width=0.75\textwidth]{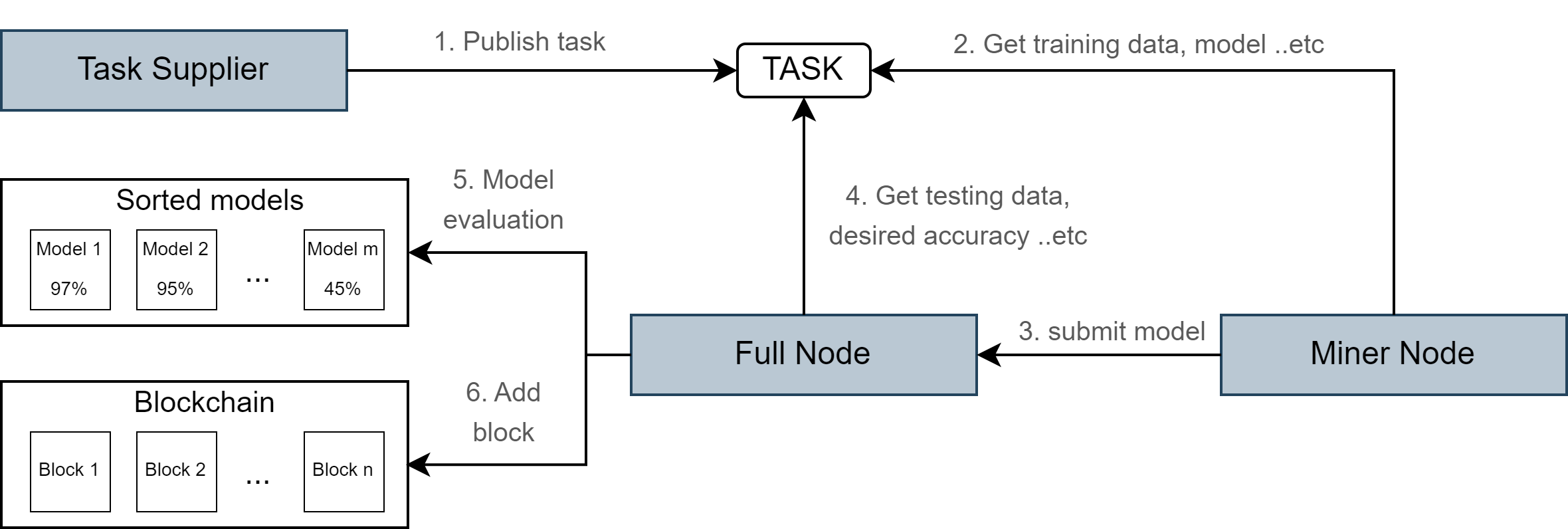}
    \caption{Proof of deep learning (PoDL) Workflow.}
    \label{workflow}
\end{figure*}

Most of the proposed techniques such as in \cite{b2, b3, b4, b5, b6, b7, b80, b81} rely on three main participants for achieving consensus in PoDL, which are:

\begin{itemize}
  \item \textbf{Miners}: The nodes within the network contend with one another to append new blocks onto the chain. In the original PoW consensus mechanism, miners compete to resolve a computationally difficult hash puzzle. In the PoDL consensus mechanism, miners contend to submit trained models as proof of work. For this reason, some researchers refer to them as "trainers" (e.g., \cite{b2}).
  
  \item \textbf{Full nodes}: These nodes uphold the blockchain and authenticate the work of miners by scrutinizing their performance (e.g., accuracy, precision, recall, etc. of the submitted models). Some researchers such as in \cite{b2} call these nodes "validators" because they handle the validation of the submitted models.

  \item \textbf{Task/Data publishers}: These are the nodes in the network that are responsible for publishing machine learning training tasks to the network. For this reason, some researchers refer to them as "suppliers" (e.g., \cite{b2}) while others call them "model requesters" such as in \cite{b3}. These tasks include the training data, testing data, metrics, the minimum threshold for each metric, the model architecture to be trained, and any other hyper-parameters.

\end{itemize}

PoDL was originally proposed in \cite{b3} as a method for training deep learning models on decentralized networks. The workflow of PoDL can be divided into seven phases:

\begin{enumerate}

\item \textbf{Training data release}: The data publisher releases the training data, hyperparameters, pre-trained model (if any), and any other necessary requirements for conducting the training to all miners. 

\item \textbf{Training}: The miners train the released model (if there is any) on the released training data.

\item \textbf{Block header submission}: Once the miners finish the training, they compute the value of the hash of the trained parameters for the model and then send the header of the block to the full nodes.

\item \textbf{Testing data release}: The data publisher releases the test data (or part of it) to all nodes.

\item \textbf{Assessing model performance}: The nodes then utilize the test data to calculate the trained model's accuracy.

\item \textbf{Model submission}: Once the nodes have finished calculating the accuracy; the trained model, together with the block, is submitted for validation.

\item \textbf{Block selection}: The validators then assess the submitted models' accuracy and sort the models based on their accuracy. Once that is done, the validators then accept the first model with the highest accuracy. In the event of a tie, the first block sent will be chosen, or some researchers prefer to choose the smallest model size, as in \cite{b2}.

\end{enumerate}

Figure \ref{workflow} illustrates an overview of the workflow described above from the task generation by the task publisher till the block gets added to the blockchain by the full nodes.

\subsection{Design of Reward}
\label{reward-design}
In the original PoW-based blockchain, the miners are rewarded with transaction fees and block rewards for successfully solving a computationally difficult puzzle \cite{incentive-analysis}.

In PoDL, several approaches for reward distributions have been proposed. For example, in \cite{b2}, the authors proposed a PoDL blockchain where miners are rewarded by task publishers for training the best-performing model for a given task. The PoDL blockchain validators are compensated equally with a transaction fee provided by the task publisher as well as new WekaCoins. The proposed method also prevents the task publisher from participating in the competition by training on the test data. This is because the task publisher would have to pay for both itself and the validators, which is infeasible from the task publisher's point of view.

In \cite{b80}, the authors designed a reward mechanism that incentivizes miners to train models for task publishers. In this system, task publishers pay miners to train a model on their data. The miners are rewarded with the publisher's payment. No new coins are generated in this process, which prevents the task publisher from participating in the competition.

\subsection{Task initiation process}
Task initiation is the process of requesting a model to be trained on a specific dataset. In \cite{b2}, researchers proposed a special transaction called a ``task publication transaction" while in \cite{b81} the authors call it ``data transaction". In this transaction, the task publisher proposes a challenge to train a model. This transaction contains the information required to complete the task, such as the dataset, the desired performance, and the reward for completing the task.

Once the task transaction is ready to be instantiated, it is important to digitally sign both the transaction and the data \cite{b2}. This will prevent any malicious intermediary in the network from manipulating the data to make other nodes perform maliciously. For example, an attacker could modify the data in a way that causes other nodes to train on incorrect data. This would waste the resources of the other nodes and increase the attacker's chance of winning the next block.

\subsection{Model submission process}
Once a miner node completes the training process, to inform the network that a trained model is ready, the authors in \cite{b2} used a special transaction called ``model transaction" wherein the trainer puts forth a solution for a specific task. The model transaction is digitally signed to prevent any manipulation by intermediaries. This is similar to the task initiation transaction, which is also digitally signed. The digital signatures ensure that the transactions have not been tampered with, and that they are created by an authorized sender. This contributes to the fairness and integrity of the competition data\cite{dig-sig}.

\subsection{Validation and Block Acceptance Criteria}
In the original PoW-based blockchain, the first miner that proposes a valid block will get its block appended to the blockchain \cite{b0}. The work acceptance criterion in PoDL is dependent on the metrics specified by the task publisher. Once the task publisher releases the testing data, the miners then send the calculated metrics, together with the model, to the full nodes. The latter check the metrics and sort all of the submitted models by accuracy. The most accurate model is then accepted by the full nodes. Furthermore, to ensure that no node steals a trained model from other nodes during the model submission process, all nodes are first required to submit the header of the block that includes the hash of the model parameters as well as the result in some cases. This ensures that even if a node steals a trained model from another, when it submits the stolen model, the validator will check the submitted header as well as the model, by passing the test set to the model and checking if they match their claimed results in the header of the block, as well as the model parameters' hash. If it is not matched or if the miner submits a trained model without submitting the block header, the block will be rejected. Additionally, since the test data is published after the block header is submitted by the miners, the resulting model trained on the test data by any node will result in an invalid claim between the actual model results and the claimed ones in the block header \cite{b8}.

In \cite{b5}, an extra layer called the secure mapping layer (SML) was used to prevent model theft. This layer allows nodes to share their models in the network without having to submit the block header beforehand as mentioned above, since the SML is treated as part of the model. The SML takes the input data concatenated with the current and the previous block hashes as input to the model. Consequently, the input features of the model are dependent on miner information due to the coinbase. Thus, the act of model stealing can negatively impact model performance, as the node attempting to steal the model would introduce unseen features during training, leading to poor performance.

\subsection{Short time training handling and incremental training}

In many blockchain systems, the pace of block creation is fixed. For example, every 10 minutes, the Bitcoin network generates a new block \cite{b0}. However, complex deep learning models can take days or even weeks to train, especially if the model and the data will be trained on is large or the desired accuracy is high. As a result, miners may not be able to complete more than a single epoch of training during this time period. This can lead to low-accuracy models, as miners are not able to train their models on enough data or for longer periods of time.

In order to address this issue, researchers have proposed a number of solutions. One approach proposed in \cite{b3} and adopted in \cite{b8} is to allow miners to train their models over multiple blocks. In this approach, the task publisher does not gather the trained model unless the accuracy of the model does not vary throughout a number of blocks. This ensures that miners have enough time to train their models on enough data to achieve the desired accuracy. However, there is a potential for cheating in this approach. Since the task publisher will expose the test data to mine the first block, it is possible for a malicious node to deceive by training the model on data from the test set and achieve the highest accuracy in the next block. To prevent this, it is important for the task publisher to publish a fresh and new test set for each block. This allows validators to validate the work over each block. Miners keep training their models on the same training set, but they are evaluated on a fresh test set each time. This helps to ensure that miners are not cheating by training their models to the current testing data.

In \cite{b6}, the authors used a similar approach, but instead of blindly relying on the accuracy to settle and not change for a certain number of blocks, they used short-term target accuracy and desired accuracy. The desired accuracy takes longer to achieve than the block generation rate, and it is achieved by incremental improvements based on short-term goals. These short-term targets can be achieved within a pre-defined time window that is less than or equal to the block generation rate. The short-term targets are based on the results of the last task. For example, if the last task achieved 90\% accuracy while the desired accuracy is 95\%, the next short target might be 92\%.

Other researchers have made assumptions in their designs that prioritize the block generation rate over the model training. For example, in \cite{b4, b7}, the authors prioritize the block generation rate by assuming that the model training can be interrupted. This assumption is based on the fact that many machine learning models are trained using gradient-based optimization, which can be interrupted without losing much progress as long as the miners save the best checkpoint.

\subsection{Double Spending Prevention}

Double spending happens when an amount of money is spent more than once, and the transaction is successfully processed for each spending. This is done by transferring a certain amount of coins (in the case of cryptocurrencies) to user A, and then transferring the same amount to user B \cite{bc-ds2, bc-ds1}. Double spending attacks are deterred by the block acceptance policy mentioned previously, where only the block created by the trainer that trained the highest accurate model is accepted by the validators. The authors in \cite{b3} used the same techniques and found that double spending is unlikely to occur when the majority (i.e., 51\%) of full nodes are controlled or owned by a single entity. This is because the global optimum is not known in advance, so when only the most accurate models are accepted, it becomes difficult to improve upon the best-performing model. Their claim is supported by the fact that the model's accuracy is dependent on randomly chosen hyperparameters and initial weights.

In DLchain \cite{b6}, an attacker can launch a double spending attack by forking the blockchain several blocks behind and then building a longer chain above that forked block. However, this attack is unlikely to be successful because the honest miners will be able to build a longer chain on top of the main chain faster than the attacker can; this is similar to the analysis conducted in the original Bitcoin paper \cite{b0}. As a consequence, the attacker will never be able to successfully double-spend any transactions.

The authors in \cite{b2, b4, b5, b80} did not consider a double spending attack that could be carried out by utilizing the test data released by the task publisher. Since the test data is released to the network by the task publisher to allow the validators to validate the results, a malicious node could perform a double-spending attack via training a model on data from the test set. The number of examples in the test set is always substantially less than the training data, so the computational resources needed to train the model on it are negligible. If the attacker can create a fork in the blockchain, the block generation on the fork chain will be faster than the honest miners who generate blocks on the primary chain. Furthermore, if the intruder controls the majority of the network's nodes, this becomes critical. This kind of attack is also possible even if the training process is carried out over multiple blocks. In this case, the attacker might only train on a subset of the test data, to ensure that the test result is higher than the best-performing model. To prevent such an attack from happening, the authors in \cite{b3} mandated that all models submitted to the competition must be reproducible. This means that in order to verify the block, any full node has to be willing to retrain the winning model on the training set.

\subsection{Blockchain verification}

In the original blockchain proposed by Nakamoto \cite{b0}, the blockchain can be validated by verifying the proof of work by recalculating the hash of each block.

In PoDL, to check the validity of the entire blockchain (e.g., by the newcomer full nodes), full nodes require access to the trained model as well as the data to validate the blockchain. This is done by checking the blockchain block by block. For each block, the node checks the claimed accuracy, the trained model by the winner miner, and the generated block. To make the verification process easy for any new node, it is necessary for the task publisher and full nodes to upload the required data and models so that new full nodes can download them. This approach was adopted by several authors such as in \cite{b4, b7}.

In addition to the above criteria, in the approach proposed in \cite{b3}, the miners are required to provide all the parameters and configurations used, such as hyperparameters, initial weights, number of epochs, etc. This is to give the full nodes the flexibility and ability to verify their work by reproducing and retraining the model.

\subsection{Model size handling} 
Deep learning models can vary in size from a few kilobytes to several gigabytes. Storing these model parameters can be storage-consuming. However, there are a variety of methods that can be used to minimize the total storage required such as:

\begin{itemize}
    \item \textbf{Model compression}: Many techniques have been used to minimize the trained model size without impacting the accuracy of the models such as in \cite{dl-comp1, dl-comp2, dl-comp3, dl-comp4, dl-comp5, dl-comp6}.

    \item \textbf{Limiting the model size}: The miners or task publishers are limited to a certain model size (e.g. 10MB/model in \cite{dl-comp1, dl-comp2}). This can help to reduce the overall storage needed. 

    \item \textbf{Removing low-performing models}: This can be done if the model training process is done over multiple blocks. If a model is not performing well, it can be removed without jeopardizing the integrity of the blockchain. This is because the tamper-proofing of the blockchain is guaranteed by the high-accuracy models. This approach was used in \cite{b3, b4, b7}.
\end{itemize}

\subsection{Data privacy}
According to \cite{data-privacy}, individuals have the right to regulate the collection, utilization, and distribution of their personal information, which is also known as data privacy, data protection, or information privacy.

The training and testing data published by the task publisher may contain sensitive information. The authors in \cite{b5} proposed a solution to this problem by using data encryption. They utilized two forms of ciphertexts which are: 1) inner-product functional encryption (IPFE) \cite{IPFE} and 2) IPFE with function-hiding (IPFE-FH) \cite{IPFE-FH}. It serves to secure the privacy of data and prevent it from being abused.

\subsection{Selfish Publisher Attack}

This kind of attack was first discussed in \cite{b6}. Since the identities of the miners and the task publisher are not known, there is nothing to prevent the task publisher from participating as a miner node.

This attack is common when the training process is done over multiple blocks and the desired accuracy is achieved by setting short-term goals for each block. The attack is carried out when the task publisher has already pretrained a model in order to win the block reward. The publisher can carry out the attack in two different ways:

\begin{enumerate}
    \item Set the short-term goal to be easy to win the reward using its pre-trained model.
    \item Set the short-term goal to a high value in such a way the training takes a long time to reach the desired target, and the publisher can generate valid blocks with its pre-trained model.
\end{enumerate}

For the first attack, the authors in \cite{b6} solved it by discarding the blocks. This is because too short block intervals of block generation indicate that the difficulty of the task is low.

The second attack is very difficult to carry out after reaching a certain accuracy. This is because there is competition between nodes, and the attacker must produce a well-trained model that requires resources to achieve. As a result, it may be infeasible to carry out the attack.

If task publishers pay nothing or only a small amount of money for the submitted tasks and there is a block reward and transaction fees, this type of attack may be feasible. However, section \ref{reward-design} discusses how the reward structure can be designed to prevent such attacks.

\section{PoDL Challenges}
\label{sec:challanges}

The authors in \cite{challenges} studied the challenges of proof-of-useful-work (PoUW) systems. These challenges can be summarized and reflected in PoDL as follows:

\begin{itemize}

    \item \textbf{Block sensitivity and non-reusability}: Without these properties, it would be possible to pre-calculate future blocks using existing proofs-of-work. To retain block sensitivity, the hash from the preceding block must be used in the current block. To retain non-reusability, it is necessary to bind the validity of a PoW to the block it validates.
    
    \item \textbf{Adjustable problem hardness}: The hardness of the given problems can be adjusted to meet the target difficulty, ensuring a consistent block generation rate. This is crucial in PoDL, as having a large dataset with complex models, the best-performing node -in terms of its computational power- may not be able to complete a short-term goal, or even complete a single epoch.

    \item \textbf{Fast verification}: A full node needs to have the capability to quickly verify the validity of a block proposed by the miners. This is done by checking the proof-of-work done by the miners. In the case of PoDL, if the full nodes are going to retrain the model to verify the work done, it might not be a good idea. This is because having a complex model and large data may lead to slowing down the verification process, which in turn allows attackers to perform spam attacks on the network.

    \item \textbf{Problem is parallelizable}: miners should utilize their full computational power to propose the next block. In PoDL, this can be done by utilizing vectorization to optimize and speed up the training process.

\end{itemize}
   
\section{Discussion and Future Directions}
\label{sec:disc}

Even though PoDL is a growing field of interest in the research community, and not that much work has been done in that field, it's promising as it, on one hand, can maintain the immutability and security of the blockchain network, while on the other hand, it makes the process more efficient and helps in the development of other fields such as AI.

PoDL enables researchers and developers to submit a training task and get the results without the need for manually setting up a training environment on the cloud or internal infrastructure. It benefits both parties the researcher and developer on one hand and the miners and the blockchain on the other hand, by providing a trained model to researchers/developers and compensating miners for their work, thereby aiding the blockchain network's security. 
Furthermore, researchers and developers can conduct numerous experiments with various model architectures and hyperparameters utilizing the blockchain network. Different nodes can execute a diverse set of experiments, saving individuals time and resources.

Despite the progress that has been made, and the advantages that PoDL brings, there is still a lot of work to be done to improve the current systems. Here are some of the aspects that need further investigation and research:

\begin{itemize}
    \item \textbf{Data privacy}: As we mentioned earlier, data may contain sensitive information. However, not much work has been done in this area to protect the submitted data by the task publisher.

    \item \textbf{Double spending attack}: As we mentioned in our work, there is still no fully mature system that prevents such attacks from happening. As mentioned earlier, an attacker could use the test data and train the model on it to perform a double spending attack by creating a fork in the blockchain.

    \item \textbf{Continuous task suppliers}: It is critical to have a continuous stream of tasks that can be published to the miners to train models. If there are no tasks, the network will be jeopardized.

    \item \textbf{ASIC hardware}: The Application-Specific Integrated Circuit (ASIC) hardware is a chip created for a specific task. In cryptocurrency mining, ASICs are preferred over general-purpose CPUs and GPUs as they offer better efficiency. However, there is limited research on the impact of ASIC hardware designed for Bitcoin on PoDL. It may be necessary to use new hardware like Tensor Processing Units (TPUs) for PoDL \cite{b3}. 

    \item \textbf{Collusion prevention}: Strategies must be applied to prevent malicious collaboration among the primary participants of the network - validators, trainers, and suppliers.

\end{itemize}

\section{Conclusion}
\label{sec:conc}
In this work, we reviewed the latest approaches for PoDL as an example of PoUW. We have discussed the workflow of different PoDL algorithms and their advantages and disadvantages, the network participants, and their interactions. We have also discussed the challenges of implementing PoDL and future research directions.

PoDL is a promising new consensus mechanism that has the potential to shift the resources used in the Proof of Work (PoW) to be more useful. However, some challenges must still be overcome before PoDL can be widely used. One challenge is that PoDL requires the model publishers to continuously provide the network with new tasks. Another challenge is data privacy.

Despite these challenges, PoDL is a promising new technology, and we believe that PoDL has the potential to make PoW-based blockchains more efficient.

\end{document}